\documentclass[%
aps,
pre,
superscriptaddress,
a4paper,
12pt,
longbibliography,
reprint,
notitlepage,
floatfix
]{revtex4-2}
\usepackage[english]{babel}
\usepackage{amssymb,amsmath,stmaryrd,array}

\usepackage{graphicx}
\usepackage{subfig}

\makeatletter

\newcommand{\sca}[2]{\ensuremath{\bigl({#1}\cdot{#2}\bigr)}}

\newcommand{\hcnj}[1]{{#1}^{\dagger}}

\newcommand{\prt}[1]{\partial_{#1}}

\newcommand{\ddiv}{\mathop{\rm div}\nolimits}

\newcommand{\mum}{$\mu$m}
 \newcommand{\bs}[1]{\boldsymbol{#1}}
 \newcommand{\vc}[1]{\mathbf{#1}}
 
 \newcommand{\uvc}[1]{\hat{\mathbf{#1}}}
 
 \newcommand{\ind}[1]{\mathrm{#1}}

\newcommand{\dd}{\mathrm{d}}

\makeatother

\begin{document}
\title{Hysteresis and Fr\'eedericksz thresholds for twisted states in chiral nematic liquid crystals: Minimum-energy path approach}

\author{Semen~S.~Tenishchev}
\email{tenischev.semen@gmail.com}
\affiliation{ITMO University, Kronverkskiy, 49, 197101, St. Petersburg, Russia}
\affiliation{St.\,Petersburg State University, 199034, St.\,Petersburg, Russia}

\author{Ivan~M.~Tambovtcev}
\email[Email address: ]{imtambovtcev@gmail.com}

\affiliation{St.\,Petersburg State University, 199034, St.\,Petersburg, Russia}

\author{Alexei~D.~Kiselev}
\email[Email address: ]{alexei.d.kiselev@gmail.com}
\affiliation{ITMO University, Kronverkskiy, 49, 197101, St. Petersburg, Russia}
\affiliation{St.\,Petersburg State University, 199034, St.\,Petersburg, Russia}

\author{Valery~M.~Uzdin}
\email[Email address: ]{v\_uzdin@mail.ru}
\affiliation{St.\,Petersburg State University, 199034, St.\,Petersburg, Russia}
\affiliation{ITMO University, Kronverkskiy, 49, 197101, St. Petersburg, Russia}

\begin{abstract}
  We study minimum-energy pathways (MEPs)
  between the branches of metastable helical structures
  in chiral nematic liquid crystals (CNLCs)
  subjected to the electric field
  applied across the cell.
  By performing stability analysis
  we have found that,
  for the branches with non-vanishing half-turn number,
  the threshold (critical) voltage
  of the Fr\'eedericksz transition
  is an increasing function of the free twisting wave number.
  The curves for the threshold voltage depend on the elastic anisotropy
  and determine the zero-field critical free twisting number where
  the director out-of-plane fluctuations destabilize the CNLC helix.
  For each MEP passing through a first order saddle point
  we have computed the energy barrier as the energy difference between
  the saddle-point and the initial structures at different values of
  the applied field.
  In our calculations, where the initial approximation for
  a MEP at the next step was determined by the MEP obtained at the
  previous step, the electric field dependence of the energy
  barrier is found to exhibit the hysteresis.
  This is the hysteresis of electrically driven transition of
  the saddle-point configuration between
  the planar and the tilted structures involving out-of-plane director deformations.
  It turned out that, by contrast to the second-order
  Fr\'eedericksz  transition,
  this transition is first order
  and we have studied how it depends
  on the zenithal anchoring energy
  strength.% and the elastic anisotropy.
\end{abstract}

% \pacs{%
% 03.65.Vf, 07.60.La,
% 42.25.Hz, 42.25.Ja,
% 78.20.Jq, 42.70.Df
% }
% \keywords{%
% twisting transitions;
% chiral liquid crystals; anchoring breaking;
% }

\maketitle

%%%%%%%%%%%%%%%%%%%%
\section{Introduction}
\label{sec:intro}
%%%%%%%%%%%%%%%%%%%

Self-organized soft helical superstructures
appear as a manifestation of the chirality caused by
the broken mirror symmetry
in certain liquids with long-range orientational order
known as \textit{liquid crystals} (LCs)
where the molecules tend to align along
a unit vector $\uvc{n}(\vc{r})$ called
the LC \textit{director}
specifying
the locally averaged direction of the LC molecules at a point
$\vc{r}$~\cite{Gennes:bk:1993,Oswald:bk:2005}.
More specifically,
in chiral liquid crystals,
the presence of chiral molecules
(molecules with no mirror plane)
gives rise to equilibrium structures
forming helical twisting patterns
where the director rotates about a twist (helical) axis.

These supramolecular helical structures
are behind a unique combination of photonic
properties of chiral nematic liquid crystals
also known
as the \textit{cholesteric liquid crystals} (CLCs).
Tunability of such structures
underlies most of
the fascinating device applications of CLCs~\cite{Yang:bk:2006,Obayya:bk:2016}.
So controllable manipulation of the CLC helical
superstructures
presents a challenging problem which is of vital importance for
both fundamental and technological
reasons~\cite{Bisoyi:advmat:2018,Zola:inbk:2018,Ryabchun:adom.2018,Balamurugan:rfp:2016}.

The continuum theory of CLCs
is formulated in terms of
the Frank-Oseen free energy functional 
\begin{align}
  \label{eq:frank}
  &
    F_{\ind{el}}[\vc{n}]=\int_V f_{\ind{el}}\,\dd v,\quad
    f_{\ind{el}}=\frac{1}{2} \Bigl\{ K_1 ({\nabla}\cdot\vc{n})^2
    \notag
  \\
  &
    +K_2
    \left[ \vc{n}\cdot{\nabla\times\vc{n}}+q_0 \right]^2
+K_3\,[\vc{n}\times(\nabla\times\vc{n})]^2
    \notag\\
  &
 -K_{24} \ddiv\left(
\vc{n}\ddiv\vc{n}+ \vc{n}\times(\nabla\times\vc{n}) \right)
\Bigr\}\,, 
\end{align}
where
$f_{\ind{el}}$ is the elastic free energy density;
$K_1$, $K_2$, $K_3$ and $K_{24} $ are the splay, twist, bend
and saddle-splay Frank elastic constants,respectively.
The bulk free energy~\eqref{eq:frank} contains a chiral term proportional
to the parameter $q_0$, 
the free twist wave number or the free twisting number,
giving the pitch $P_0=2\pi/|q_0|$ of equilibrium helical structures
in unbounded CLCs.

An efficient method widely used to prepare
CLCs is doping nematic LC mixtures
with chiral additives that induce a helical structure~\cite{Oswald:bk:2005,Balamurugan:rfp:2016}.
For photosensitive chiral dopants (photoswitches),
their helical twisting power and thus
the CLC equilibrium helix pitch $P_0$
can be controlled by light
through photoinduced changes in 
chiral molecular switch conformation that influence
the LC's helical twisting
power~\cite{Vinograd:mclc:1990,Delden:chem:2003,Eelkema:lc:2011,Katsonis:jmatchem:2012,Kiselev:pre:2014,Zheng:nature:2016,Bisoyi:anie:2015,Huang:marc:2017}.
Phototunability of the helix pitch
leads to a variety of technologically promising effects such as the
phototunable selective reflection,
i.e. a light-induced change in the spectral position of the bandgap~\cite{Kurihara:apl:1998,White:jap:2010,Kosa:nature:2012,Vernon:optexp:2013}.

% An important point is that, director configurations in the planar CLC cells
%  are strongly affected by the
% anchoring conditions at the substrates.
% These conditions break the translational symmetry along the twisting
% axis and, in general, the helical form of the director field will be distorted.
% Nevertheless, when the anchoring conditions are planar and
% out-of-plane deviations of the director are suppressed, it might
% be expected that the configurations still have the form of the
% ideal helical structure~\eqref{eq:clc_helix}.
% But, by contrast with  the case of
% unbounded CLCs, the helix twist wave number $q$ will now differ
% from $q_0$.

% A mismatch between 
% the twist imposed by the boundary conditions
% and the equilibrium pitch $P_0$ may produce two
% metastable twist states that are degenerate in energy and can be
% switched either way by applying an electric
% field~\cite{Berrem:jap:1981}.

An important point is that multiple metastable twist states in CLC cells generally appear
as a result of competing influences of the bulk and the surface
contributions to the free energy
leading to frustration~\cite{Kamien:jcmp:2000,Oswald:bk:2005}
and giving rise to multiple local minima
of the energy~\cite{Mottram:cmt:1997}.
Properties of the metastable helical structures are determined by 
the anchoring energy and the free twisting number $q_0$.
The latter plays the role of the governing parameter
for the
\textit{pitch transitions}
theoretically studied in
Refs.~\cite{Bel:eng:2000,Bel:eng:2003,Palto:eng:2002,Kiselev:pre-1:2005,Kiselev:pre:2014,McKay:epje:2012,Lelidis:pre:2013,Barbero:jmolliq:2018,Kiselev:pre:2014}.
These transitions occur 
between different branches of metastable
states and, in particular, manifest themselves in
a jump-like temperature dependence of selective
light transmission 
spectra~\cite{Zink:jetp:1997,Gandhi:pre:1998,Zink:1999,Yoon:lc:2006}.

Geometrically, such transitions
are related to
the free energy pathways
connecting pairs of metastable helical states
which are local minima (minimizers)
of the free energy landscape viewed as a multidimensional
free energy
surface~\cite{Bessarab:cpc:2015}.
The key elements associated with the transitions
are the \textit{minimum energy paths} (MEPs) between the initial and
final states on the free energy surface.

Every point on such a pathway
is a free  energy minimum in all but a certain direction in the
configuration space of CLC director structures. 
It represents a path with the maximal statistical weight
and provides a scenario of the most probable transition between
the states.

Maximum along
the MEP determines the \textit{transition state} which is a saddle
point on the free energy surface. 
 The energy barrier separating the states can be found as
the difference between the saddle point energy and the energy of the
initial state.
Such energy barriers are used
to assess
the effect of thermally activated transitions within
the framework of the rate
theory~\cite{Hanggi:rmp:1990,Coffey:advchem:2007}.
The barrier heights
combined with
the Arrhenius formula were
also employed
to estimate
the rate of transitions between metastable twist states
in LC cells with strong anchoring conditions~\cite{Goldbart:prl:1990,Goldbart:mclc:1991}.
The two-dimensional Landau-de Gennes theory
was used to study the MEPs of the planar bistable LC device
depending on the anchoring conditions~\cite{Kusum:softmat:2015}.

In Ref.~\cite{Ivanov:pre:2016,Kiselev:pre:2019}
the approach based on MEPs
was applied to
explore
field-induced distortions
of the free energy landscape
in CLC cells
subjected to external (magnetic or electric) fields.
These distortions
are known to lead to a variety of
field-induced orientational effects
such as
the Fr\'eedericksz and unwinding transitions.
When
the electric field is applied across the CLC cell,
these effects crucially depend on
a number of factors such as
the cell thickness $L$, the pitch $P_0$,
the applied voltage $U$, the anchoring conditions,
elastic and dielectric properties
of the CLC
material~\cite{Meyer:apl:1968,Berrem:jap:1981,Becker:jap:1985,Schiller:phtrans:1990,Hirning:jap:1991,Smalyukh:pre:2005,Choi:advmat:2009,Valkov:pre:2013}.

This geometry where
the CLC cell with planar 
anchoring conditions is subjected to
the field applied along the twisting axis
of the vertically aligned planar helical structure
will be our primary concern.
In this paper
we shall use
the geodesic nudged elastic band (GNEB) method~\cite{Bessarab:cpc:2015} to
calculate MEPs and the energy barriers
for the transitions between
metastable twisted states that belong to
the branches with different half-turn numbers.
As opposed to the case
of unwinding transition
with negligibly small free twisting number
previously studied in Ref.~\cite{Kiselev:pre:2019},
it turned out that
the half-turn numbers of metastable helix state branches
and the elastic anisotropy
are the governing factors for
$q_0$-dependence of the critical field
of the Fr\'eedericksz transition.
As a consequence
the Fr\'eedericksz thresholds
for the helical structures
involved in the transitions are generally different
and
we shall study
how the free energy barriers
of such transitions
depend on
the electric field and the anchoring conditions. 

The layout of the paper is as follows.
General relations
that determine the characteristics of
the helical structures in CLC cells are given
in Sec.~\ref{sec:theory}.
The the Fr\'eedericksz thresholds
are computed by performing
stability analysis in Sec.~\ref{sec:stability}.
Then in Sec.~\ref{sec:results} we outline the
numerical procedure that we use to
compute MEPs and the energy barriers.
Finally, in Sec.~\ref{sec:conclusion}
we discuss our results and
make some concluding remarks.

%%%%%%%%%%%%%%%%
\section{Free energy}
\label{sec:theory}
%%%%%%%%%%%%%%%%

We consider
the planar confining geometry
of a CLC cell of thickness $L$,
where CLC is placed
between two parallel bounding plates:
$z=-L/2$ (lower substrate) and $z=L/2$ (upper substrate).
Anchoring conditions at the substrates are planar
with the preferred orientation of molecules at
the lower (upper) plate
defined by
the vector of easy orientation
$\uvc{e}_{-}$ ($\uvc{e}_{+}$)
\begin{align}
  \label{eq:clc-anch-vec}
  \uvc{e}_{\pm}=\cos\psi_{\pm}\,\uvc{x}+\sin\psi_{\pm}\,\uvc{y},
\end{align}
where a hat will indicate unit vectors.

As in Ref.~\cite{Kiselev:pre:2019},
the anchoring energy $W_{\nu}(\vc{n})$
that enters the
surface contribution to the CLC free energy functional:
\begin{align}
  &
  F[\vc{n},\vc{E}]=F_b[\vc{n},\vc{E}]+F_s[\vc{n}],
    \notag
  \\
  &
  F_s[\vc{n}]=\sum_{\nu=\pm 1}
\int_{z=\nu L/2} W_{\nu}(\vc{n})\,\dd s,
\label{eq:f-gen}
\end{align}
where $\vc{E}$ is the electric field,
will be taken in the form of Rapini-Papoular potential~\cite{Rap:1969}:
\begin{align}
&
W_{\nu}(\vc{n})=
\frac{W^{(\nu)}_{\phi}}{2}
\Bigl[1-\sca{\vc{n}}{\uvc{e}_{\nu}}^2\Bigr]_{z=\nu L/2}
\notag
\\
&
+
\frac{W^{(\nu)}_{\theta}-W^{(\nu)}_{\phi}}{2}
\sca{\vc{n}}{\uvc{z}}^2\Bigr|_{z=\nu L/2},
  \label{eq:f_s-gen}
\end{align}
where $W^{(+)}_{\phi}$ ($W^{(-)}_{\phi}$) and $W^{(+)}_{\theta}$ ($W^{(-)}_{\theta}$) are the azimuthal
and the polar anchoring strengths
at the upper (lower) substrate.

For the CLC director
\begin{align}
  \label{eq:n-angles}
\vc{n}\equiv\vc{n}(\theta,\phi) = \cos\theta (\cos\phi\,\uvc{x}+ \sin\phi\,\uvc{y}) + \sin\theta\,\uvc{z}  
\end{align}
parameterized by the \textit{tilt} and the \textit{azimuthal} angles,
$\theta=\theta(z)$ and $\phi=\phi(z)$,
the surface potential~\eqref{eq:f_s-gen}
assumes the following form:
\begin{align}
  &
  \label{eq:anchoring-pot}
  F_s[\vc{n}]/A=\sum_{\nu=\pm 1}
  \biggl[
    \frac{W^{(\nu)}_{\phi}}{2}\cos^2\theta_\nu\sin^2(\phi_\nu-\psi_\nu)+
\notag
  \\
  &
  \frac{W^{(\nu)}_{\theta}}{2}\sin^2\theta_\nu
  \biggr],
\end{align}
where
$A$ is the area of the substrates;
$\theta_\nu\equiv\theta(\nu L/2)$
and $\phi_\nu\equiv\phi(\nu L/2)$.
\begin{figure}[!htb]
   \centering
   \resizebox{90mm}{!}{\includegraphics*{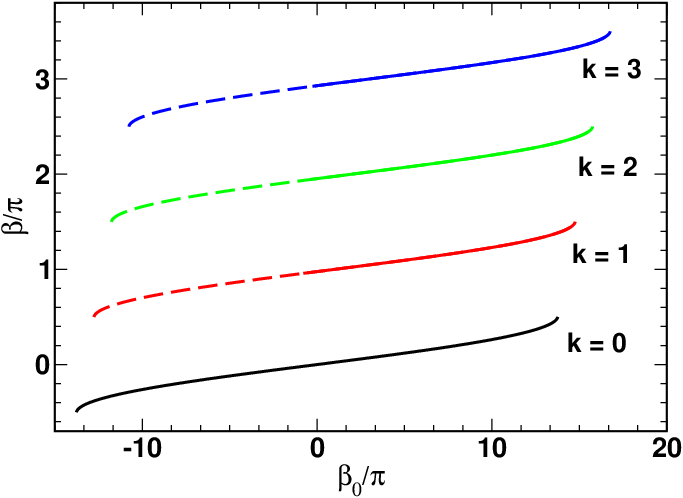}}
   \caption{Twist parameter versus
     the free twisting number $\beta_0/\pi=q_0L/\pi$ at different
     values of the half-turn
     number, $k$. Dashed lines indicate the states unstable with respect
     to out-of-plane fluctuations.
     }
\label{fig:q-q0}
\end{figure}
The bulk part of the free energy functional~\eqref{eq:f-gen}
\begin{align}
  \label{eq:bulk-energy}
  F_b[\vc{n},\vc{E}]=F_{\ind{el}}[\vc{n}]+F_{\ind{E}}[\vc{n},\vc{E}]
\end{align}
is a sum of the elastic energy
$F_{\ind{el}}[\vc{n}]$ given by Eq.~\eqref{eq:frank}
and the energy of interaction between
the electric field $\vc{E}$ and CLC molecules,
$F_{\ind{E}}[\vc{n},\vc{E}]$.
After substituting the CLC director~\eqref{eq:n-angles},
into the elastic energy~\eqref{eq:frank} we have
\begin{align}
  \label{eq:elastic-energy}
  &
  F_{\ind{el}}[\vc{n}]/A = \frac{1}{2} \int_{-L/2}^{L/2}
  \bigl\{
    K_1(\theta)[\theta']^2 +
    \notag
  \\
  &
    K_2(\theta)\cos^2\theta\, [\phi']^2 - 2C(\theta)\phi' + K_2q_0^2
  \bigr\}\dd z,
  \\
  &
  \label{eq:Ki-eff}
  K_i(\theta) = K_{i}\cos^2\theta+K_{3}\sin^2\theta,
  \: C(\theta) = q_0K_{2}\cos^2\theta,  
\end{align}
where prime stands for derivative with respect to $z$.

We also assume that the electric field
is normal to the substrates
$\vc{E}=E_z(z)\uvc{z}$
with $E_z(z)=-V'(z)$,
where $V(z)$ is the electrostatic potential,
and the applied voltage,
$U=V(-L/2)-V(L/2)$, is fixed
So, the electrostatic part of the energy
\begin{align}
  \label{eq:el-en-gen}
  F_{\ind{E}}=-\frac{1}{2}\int_V\sca{\vc{E}}{\vc{D}}\dd v,
\end{align}
where
$\vc{D}=\bs{\epsilon}\,\vc{E}$ is the electric dispacement
and $\bs{\epsilon}$ is the dielectric tensor, 
takes the form of nonlocal functional:
\begin{align}
  \label{eq:el-en}
  F_{\ind{E}}/A=-\frac{U^2}{2 E[\theta]},
  \quad
  E[\theta]=\int_{-L/2}^{L/2} \frac{\dd z}{\epsilon_{zz}(\theta)},
\end{align}
where
$\epsilon_{zz}(\theta)=\epsilon_{\perp}+\epsilon_a \sin^2\theta$,
$\epsilon_a=\epsilon_{\parallel}-\epsilon_{\perp}$;
$\epsilon_{\perp}$ and
$\epsilon_{\parallel}$ are the dielectric constants
giving the principal values of $\bs{\epsilon}$.

In our numerical calculations,
we shall use the Frank elastic constants
typical for 5CB~\cite{Bogi:lc:2001}: 
$K_{1}=4.5$~pN, $K_{2}=3.0$~pN, $K_{3}=6.0$~pN
and restrict our analysis to the case of
the symmetric CLC cell of the thickness
$L=5$~\mum\ with
$W_{\phi}^{(\pm)}\equiv W_{\phi}=0.05$~mJ/m$^2$
and $\uvc{e}_{\pm}=\uvc{x}$
($\psi_{\pm}=0$).

\begin{figure*}[!htb]
  \centering
\subfloat[5CB]{
  %\resizebox{65mm}{!}{\includegraphics*{smectic-cone.eps}}
  \resizebox{88mm}{!}{\includegraphics*{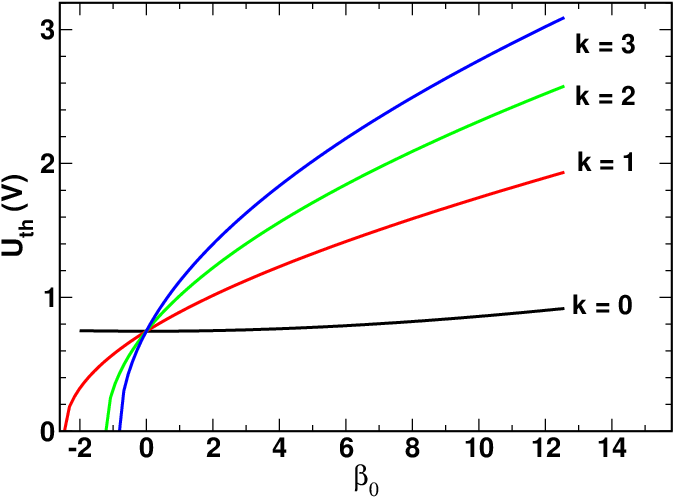}}
\label{subfig:5CB}
}
\subfloat[One-constant approximation with $K_i=K_1$ ]{
  %\resizebox{90mm}{!}{\includegraphics*{dhf-cell.eps}}
  \resizebox{92mm}{!}{\includegraphics*{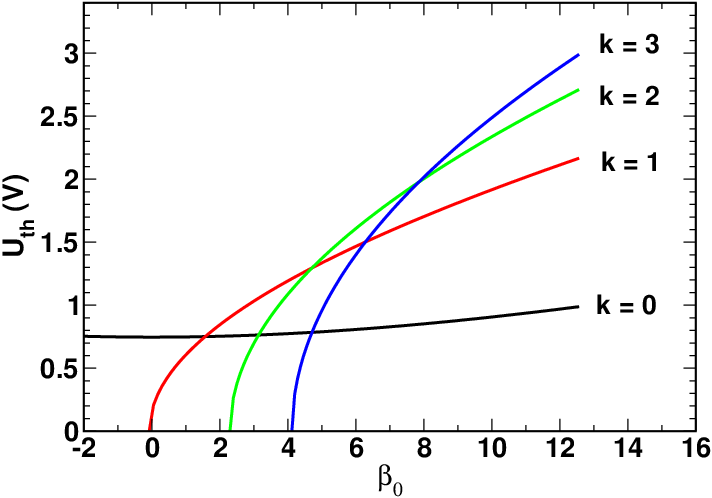}} 
\label{subfig:one-const}
}
   \caption{Critical voltage of the Fr\'eedericksz transition as a
     function of the free twisting number parameter ($\beta_0=q_0L$)
     for the branches with the half-turn numbers $0\le k \le 3$.
     The curves are computed from Eq.~\eqref{eq:stab-out}
     for the case of strong polar anchoring.
     }
\label{fig:U_th-q0}
\end{figure*}

\section{Stability and Freedericksz thresholds
of planar helical structures}
\label{sec:stability}

In this section we perform
stability analysis for the uniform standing helix state
with the twist axis normal to the substrates
known as the planar Grandjean  structure (texture)
\begin{align}
  &
  \label{eq:clc_helix}
    \vc{n}_0\equiv\vc{n}(0,u(z))=\cos u\,\uvc{x}+\sin u\,\uvc{y},
    \notag
  \\
  &
    \: u(z)=q z+\phi_0,
\end{align}
where $q= 2\pi/P$ is the \textit{helix wave number}
and $P$ the \textit{helix pitch}.
To this end, we begin with the expression for the distorted director field
\begin{equation}
  \label{eq:dir-pert}
  \vc{n}(\delta\theta,u(z)+\delta\phi)\approx\vc{n}_0+\delta\vc{n}_0,\:
\delta\vc{n}_0=\delta\phi\,\vc{n}_1+\delta\theta\,\vc{n}_2,
\end{equation}
where the angles $\delta\phi$ and $\delta\theta$ describe in-plane and out-of-plane
deviations of the director, respectively;
the vectors $\vc{n}_1$ and $\vc{n}_2$ are
\begin{equation}
  \label{eq:n12}
 \vc{n}_1=-\sin u(z)\,\uvc{x}+\cos u(z)\,\uvc{y},\quad
\vc{n}_2=\uvc{z}.
\end{equation}
The free energy of
the director field~\eqref{eq:dir-pert}
can now be expanded
up to second order
terms in the fluctuation field
$\bs{\psi}=\begin{pmatrix}\delta\phi\\ \delta\theta \end{pmatrix}$
and its derivatives
\begin{align}
  \label{eq:harm-1}
&
                      F[\vc{n}]\approx F[\vc{n}_0]+F^{(2)}[\bs{\psi}],
\\
  \label{eq:harm-2}
&
  F^{(2)}[\bs{\psi}]=\int_V f_{b}^{(2)}[\bs{\psi}]\,\dd v+
\sum_{\nu=\pm 1}
\int_{z=\nu L/2}W_{\nu}^{(2)}(\bs{\psi})\, \dd s.
\end{align}
The second order variation of the free energy $F^{(2)}[\bs{\psi}]$ is
a bilinear functional which represents the energy of the
director fluctuations written in the harmonic (Gaussian) approximation.

In what follows we shall restrict our consideration to the case
of fluctuations invariant with respect to in-plane
translations, so that $\bs{\psi}\equiv\bs{\psi}(z)$.
In this case the fluctuation energy per unit area is
\begin{align}
2F^{(2)}[\bs{\psi}]/A=
\int_{-L/2}^{L/2}
\hcnj{\bs{\psi}}\, \hat{K}
\bs{\psi}\, \dd z
+\sum_{\nu=\pm 1}
\hcnj{\bs{\psi}}\, \hat{Q}^{(\nu)}
\bs{\psi}\Bigr\vert_{z=\nu L/2},
 \label{eq:F2z}
\end{align}
where
the expressions for the operators $\hat{K}$ and $\hat{Q}^{(\nu)}$ are given by
\begin{align}
  \label{eq:op-K}
  \hat{K}=
\begin{pmatrix}
-K_2\prt{z}^2 & 0\\
0 & -K_1 \prt{z}^2+q^2 K_q-\epsilon_a E^2
\end{pmatrix},
\end{align}
\begin{align}
  \label{eq:op-Q}
 \hat{Q}^{(\nu)}=&
\nu
\begin{pmatrix}
K_2\prt{z} & 0\\
0 & K_1 \prt{z}
\end{pmatrix}
\notag\\
&
+\begin{pmatrix}W^{(\nu)}_{\phi}\cos 2 u_{\nu} & 0\\0 & W^{(\nu)}_{\theta}-W^{(\nu)}_{\phi}\sin^2u_{\nu}\end{pmatrix},
\end{align}
$E=U/L$,
$u_\nu=u(\nu L/2)-\psi_\nu$
and $K_q$ is the effective elastic constant
\begin{equation}
  \label{eq:Kq}
  K_q = K_3 - 2 K_2 (1-q_0/q).
\end{equation}
Spectrum  of $\hat{K}$ can be computed
by solving the boundary-value problem:
\begin{align}
  \label{eq:gen-EV}
  \hat{K}\bs{\psi}_{\lambda}=\lambda\bs{\psi}_{\lambda},
\\
\label{eq:gen-EV-bc}
\hat{Q}^{(\nu)}
\bs{\psi}_{\lambda} \Bigr\vert_{z=\nu L/2}=0,
\end{align}
where
$\lambda$ ($\bs{\psi}_{\lambda}$)
stands for the eigenvalue  (the eigenmode).
From Eqs.~\eqref{eq:op-K}-\eqref{eq:op-Q}
the operators $\hat{K}$ and $\hat{Q}^{(\nu)}$ are both diagonal,
so that the in-plane and out-of-plane fluctuations
can be treated separately.
An important point is that
the helical configuration~\eqref{eq:clc_helix}
is locally stable provided that all the eigenvalues are positive.

\begin{figure*}[!htb]
   \centering
\subfloat[$k=1$, $U=0$~V]{
  \resizebox{51mm}{!}{\includegraphics*{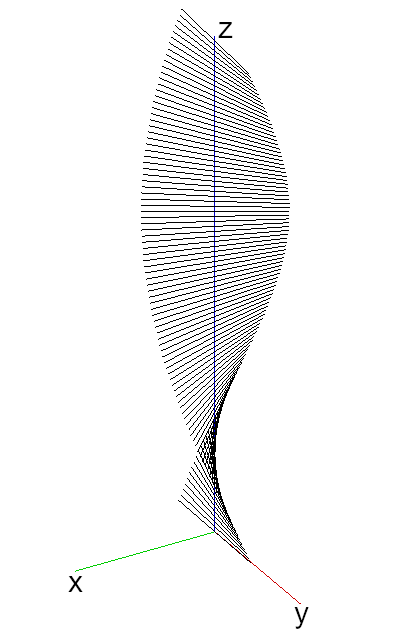}}
  \label{subfig:k1-0}
}
\subfloat[$k=1$, $U=1.4$~V]{
  \resizebox{51mm}{!}{\includegraphics*{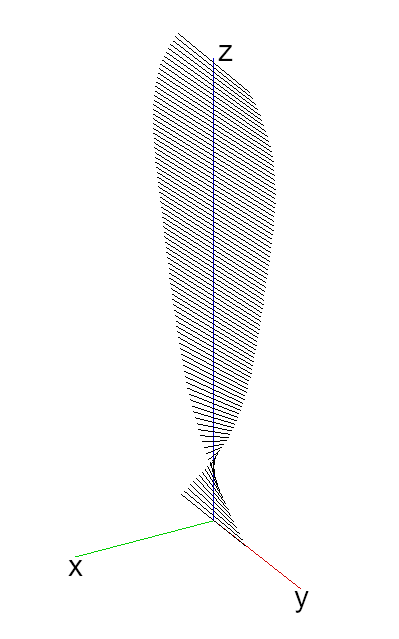}} 
\label{subfig:k1-1}
}
\subfloat[$k=2$, $U=0$~V]{
  \resizebox{51mm}{!}{\includegraphics*{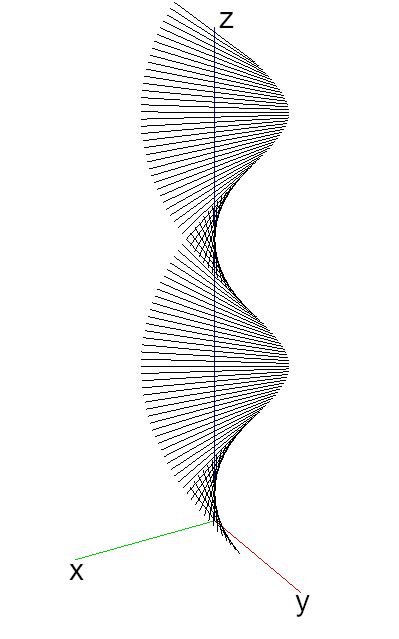}} 
\label{subfig:k2-0}
}
\caption{Metastable helical configurations:
  (a)~field free one-half-turn structure with $k=1$;
  (b)~distorted one-half-turn structure at $U=1.4 V$; and
  (c)~field free one-full-turn structure with $k=2$.
  Parameters are: $W_{\phi}^{(\pm)}\equiv W_{\phi}=0.05$~mJ/m$^2$,
  $W_{\theta}^{(\pm)}\equiv W_{\theta}=0.5$~mJ/m$^2$ and $q_0L=4.725$.
     }
\label{fig:metastable-states}
\end{figure*}

For the case of  the symmetric CLC cell,
stability analysis with respect to in-plane fluctuations
performed in Ref.~\cite{Kiselev:pre-1:2005}
shows that
there are different branches of  metastable helical states
which are labelled by
the \textit{half-turn number}, $k\in\mathbb{N}$,
and where
the twisting parameter
$  (k-1/2)\pi <q L\equiv\beta<(k+1/2)\pi$
is related to the free twisting number, $q_0$,
as follows
\begin{align}
  &
  \label{eq:beta0-beta}
    q_0 L\equiv\beta_0=\beta+(-1)^kw_{\phi}\sin\beta,
    \quad
    w_{\phi}\equiv\frac{W_{\phi}L}{2 K_2}.
\end{align}
The branches with $0 \le k\le 3$
and $w_{\phi}\approx 41.67$
($W_{\phi}=0.05$~mJ/m$^2$)
are depicted in Fig.~\ref{fig:q-q0}.

%\subsection{Out-of-plane fluctuations}
%\label{subsec:out-of-plane}

We now study stability of the helical structures
with respect to the out-of-plane fluctuations.
To this end we replace $\lambda$ with $K_1 (2/L)^2\lambda$
and rewrite
the eigenvalue problem~\eqref{eq:gen-EV}-\eqref{eq:gen-EV-bc} for
$\theta$ in the following form:
\begin{align}
&
  \label{eq:str-EV}
                \bigl[\prt{\tau}^2+(-r_q+\epsilon_aU^2/K_1)/4+\lambda\bigr]\,\theta_{\lambda}(\tau)
                = 0,
\\
&
\label{eq:w-BC}
\Bigl[\pm\prt{\tau}
\theta_{\lambda}+w^{(\pm)}_{\theta}\theta_{\lambda}
\Bigr]_{\tau=\pm 1}=0,
\\
&
  \label{eq:rq}
  r_q=(qL)^2 K_q/K_1=(r_3-2r_2)\beta^2+2r_2\beta_0\beta ,
\\
&
  \label{eq:par-w-theta}
  w^{(\nu)}_{\theta}\equiv
     \frac{(W^{(\nu)}_{\theta}-W^{(\nu)}_{\phi}\sin^2u_{\nu})L}{2K_1},
     \:
     \beta=qL,
\end{align}
where $\tau\equiv 2 z/d$, $r_i\equiv K_i/K_1$
and $\beta_0=q_0 L$.

The stability condition $\lambda>0$ can now be readily written
as follows
\begin{equation}
  \label{eq:stab-out}
  U<U_{\ind{th}}=\sqrt{(4 \lambda_{\ind{min}}+r_q )K_1/\epsilon_a},
\end{equation}
where $\lambda_{\ind{min}}$ is the lowest eigenvalue
of the problem~\eqref{eq:str-EV}-\eqref{eq:w-BC} computed at $r_q=0$
and $U=0$.
The expression on the right hand side of
the inequality~\eqref{eq:stab-out} gives the critical voltage of
the Fr\'eedericksz transition.
Above this voltage
the applied electric field makes
the vertically standing helix unstable with respect to out-of-plane director fluctuations.

When the polar anchoring is strong at both substrates, $W^{(\pm)}_{\theta}\to\infty$,
the eigenvalue $\lambda_{\ind{min}}$ is known:
\begin{equation}
  \label{eq:lmb-m-str}
  \lambda_{\ind{min}}=(\kappa_{\ind{min}})^2=\pi^2/4.
\end{equation}
Otherwise, $\kappa_{\ind{min}}$ is below $\pi/2$ and can be computed as the root
of  the transcendental equation deduced in
Appendix of Ref.~\cite{Kiselev:pre-1:2005}
\begin{equation}
  \label{eq:lambda1}
  \tan 2\kappa_{\ind{min}}=
\frac{\kappa_{\ind{min}} (w^{(+)}_{\theta}+w^{(-)}_{\theta})}{\kappa_{\ind{min}}^2-w^{(+)}_{\theta} w^{(-)}_{\theta}},
\end{equation}
where $0 <\kappa_{\ind{min}}\le \pi/2$.

The curves for the critical voltage 
plotted against the free twisting number are shown
in Fig.~\ref{fig:U_th-q0}.
The results evaluated using
the one-constant approximation
(see Fig.~\ref{subfig:one-const})
are compared with
the critical field
computed for the elastic constants
listed at the end of Sec.~\ref{sec:theory}.
It can be seen that,
at non-vanishing half-turn numbers,
in both cases
the critical field (Fr\'eedericksz threshold)
for each branch
grows with
the free twisting number $q_0$.
Given the branch with $k\ge 1$,
the value of $q_0$
at which the threshold vanishes
determines the boundary point
giving the lowest free twisting
number of locally stable structures.
When $q_0$ is below this value,
the helical states are unstable with respect out-of-plane
fluctuations.
In Fig.~\ref{fig:q-q0}
such structures are indicated by the dashed lines.
A comparison between
the curves presented in
Figs.~\ref{subfig:5CB}
and~\ref{subfig:one-const}
also shows that
the elastic anisotropy has a profound effect on
the free twisting number dependence of
the  Fr\'eedericksz threshold.
Our concluding remark in this section
is that
we have additionally estimated
the Fr\'eedericksz thresholds
numerically.
The numerical results are found to be
in excellent agreement with
the theoretical predictions.

\begin{figure*}[!htb]
   \centering
\subfloat[Director slippage-to-anchoring breaking transition.]{
  \resizebox{92mm}{!}{\includegraphics*{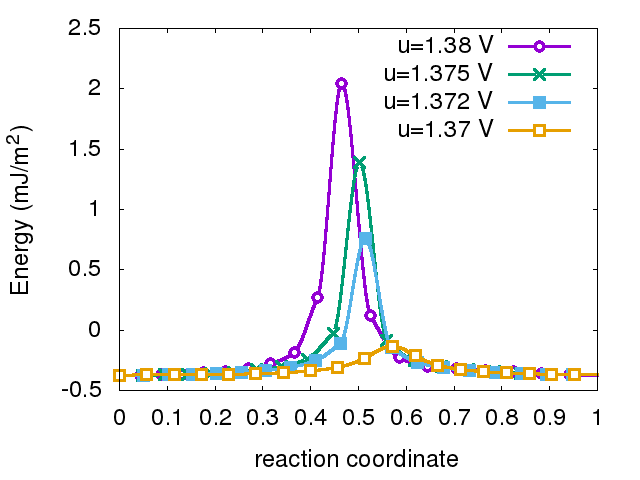}}
\label{subfig:planar-to-tilt}
}
\subfloat[Anchoring breaking-to-director slippage transition.]{
  \resizebox{92mm}{!}{\includegraphics*{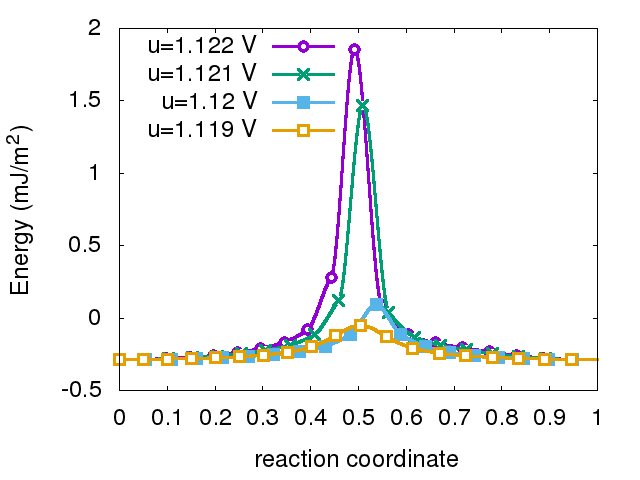}} 
\label{subfig:tilt-to-planar}
}
   \caption{Energy per unit area along the minimum energy paths at
     different voltages.
     Parameters are listed in the caption of Fig.~\ref{fig:metastable-states}.  
     }
\label{fig:soft-transition}
\end{figure*}

%%%%%%%%%%%%
\section{Numerical results}
\label{sec:results}
%%%%%%%%%%%%

In this section we present the results for
the transition between the twisted states
with
the half-turn numbers
$k=1$
(one-half-turn structure)
and $k=2$
(one-full-turn structure).
The director distributions for these
states should be solutions of the stationary point boundary
problem (the Euler-Lagrange equations
supplemented with the corresponding boundary
conditions).
In order to find the distributions
shown in Fig.~\ref{fig:metastable-states}
we have started from the initial
approximation for the director structure and then minimized
the energy using the velocity projection
algorithm~\cite{Jonsson:inbk:1998}.
The Fr\'eedericksz thresholds for the helical states
with $k=1$ and $k=2$
can be theoretically estimated
using formula~\eqref{eq:stab-out}.
So, we have the following estimates:
$U_{\ind{th}}^{(1)}\equiv U_{\ind{th}}|_{k=1}\approx 1.3$~V
and $U_{\ind{th}}^{(2)}\equiv U_{\ind{th}}|_{k=2}\approx 1.67$~V
at $\beta_0=4.725$.
In particular, as is illustrated in Fig.~\ref{subfig:k1-1},
the one-half-turn structure becomes distorted
when the applied voltage $U=1.4$~V is above $U_{\ind{th}}^{(1)}$.

Note that the above states being different in the parity of half
turns are topologically distinct.
As a result, in the strong anchoring limit, the one-half-turn helix cannot be smoothly deformed
into the one-full-turn structure
without destroying the local 
degree
of molecular ordering.
Contrastingly, in the weak anchoring
regime, these states are local minima of
the free energy surface separated by the finite energy barriers.

In this section, dependence of the energy barriers
on the applied voltage will be the subject of our primary concern.
For this purpose, we,
following our previous work~\cite{Kiselev:pre:2019},
shall use the minimum-energy path (MEP) approach.
Leaving aside technical details about the numerical procedure
described in Refs.~\cite{Ivanov:pre:2016,Kiselev:pre:2019},
we just briefly comment on the geodesic nudged elastic band
(GNEB) method to find MEPs between local minima on the curved
manifolds~\cite{Bessarab:cpc:2015}.

\begin{figure*}[!htb]
   \centering
\subfloat[$U=1.37$~V]{
  \resizebox{40mm}{!}{\includegraphics*{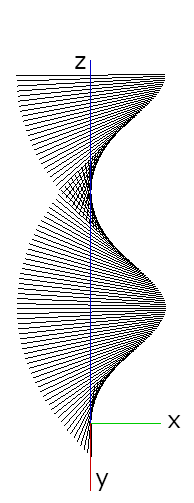}}
}
\subfloat[$U=1.372$~V]{
  \resizebox{40mm}{!}{\includegraphics*{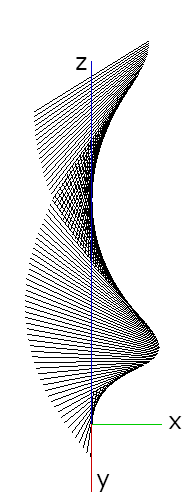}} 
}
\subfloat[$U=1.375$]{
  \resizebox{40mm}{!}{\includegraphics*{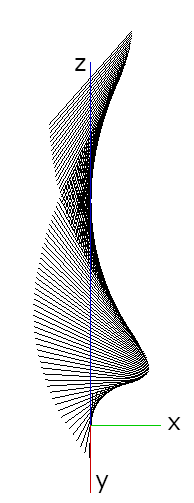}} 
}
\subfloat[$U=1.38$~V]{
  \resizebox{40mm}{!}{\includegraphics*{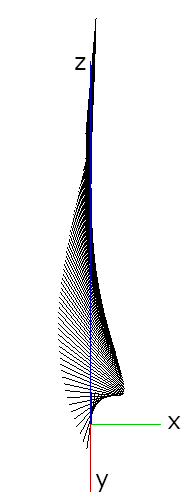}} 
}
\caption{Saddle-point structures for the MEPs with the energy curves
  depicted in Fig.~\ref{subfig:planar-to-tilt}.
  Parameters are listed in the caption of Fig.~\ref{fig:metastable-states}.  
     }
\label{fig:saddle-images}
\end{figure*}

\begin{figure}[!htb]
   \centering
   \resizebox{80mm}{!}{\includegraphics*{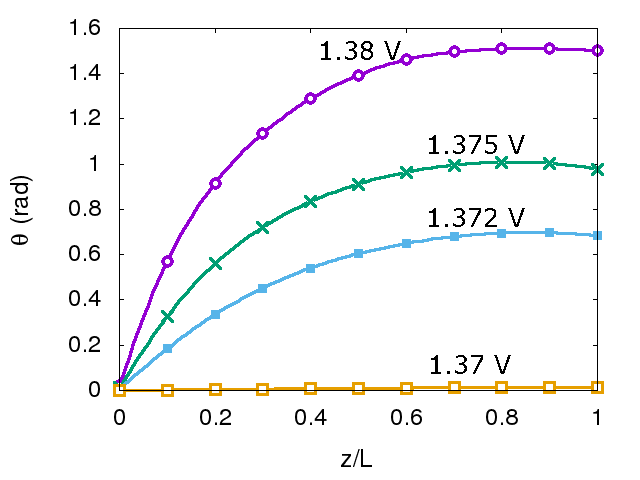}}
   \caption{Profiles of the tilt angle for the saddle point
     structures presented in Fig.~\ref{fig:saddle-images}.
     }
\label{fig:theta}
\end{figure}

This approach involves taking an initial guess of a path
between the two minima of the free energy surface
and systematically bringing that to the nearest MEP.
A path is represented by a discrete chain
of states, or ``images'', of the system, where the first and the
last image are placed at the local energy minima corresponding
to the initial and final metastable configurations.
In order to distribute the images evenly along the path, springs 
are introduced between adjacent images.
At each image, a local tangent to the path needs to be
estimated, and the force guiding the images towards the nearest
MEP is defined as the sum of the transverse component of the
energy antigradient plus the component of the spring
force along the tangent to the path. The position of intermediate images
is then adjusted so as to zero the GNEB forces.
The position of the maximum (saddle point) along
the MEP  was found using Climbing Image algorithm~\cite{Bessarab:cpc:2015}. 

\begin{figure}[!htb]
    \centering
% \subfloat[$W_{\theta}=0.5$~mJ/m$^2$]{
%   \resizebox{90mm}{!}{\includegraphics*{hysteresis.png}}
%   \label{subfig:hysteresis-05}
% }
% \subfloat[$W_{\theta}=0.1$~mJ/m$^2$]{
%   \resizebox{90mm}{!}{\includegraphics*{hysteresis_2.png}}
%   \label{subfig:hysteresis-01}
% }
    \resizebox{90mm}{!}{\includegraphics*{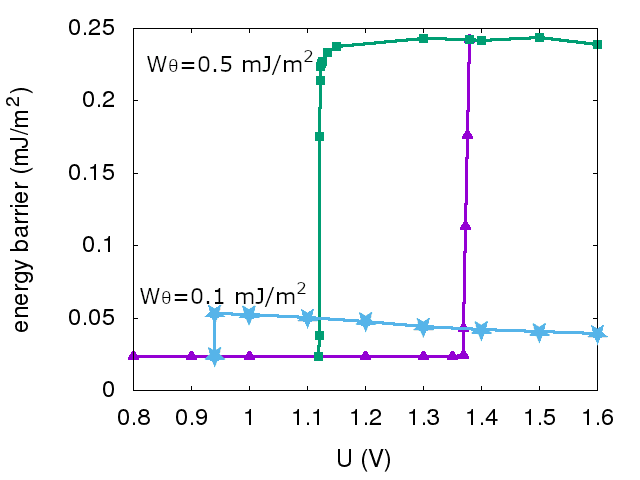}}
   \caption{Hysteresis loop in voltage dependence of the energy
     barriers computed for smoothly continued MEPs assuming the director slippage-to-anchoring breaking
     scenario (purple filled triangles)
     and the anchoring breaking-to-director slippage
     scenario (green filled squares and blue stars) at different values of
     the  polar anchoring energy strength.   
     Azimuthal anchoring energy strength is $W_{\phi}=0.05$~mJ/m$^2$.
     }
\label{fig:hysteresis}
\end{figure}

An important point is that
the MEP connecting the metastable structures
generally depends on the starting approximation for the path.
Variations in the initial approximations may produce
different MEPs.
In Ref.~\cite{Kiselev:pre:2019},
we have used different initial paths
to study two scenarios of the unwinding transition:
(a)~the \textit{director slippage transitions}
where $\theta\approx 0$;
and
(b)~the \textit{anchoring breaking transitions}
that involve tilted CLC states.

In our calculations,
similar to the director slippage scenario,
the initial (director slippage) approximation for the MEP
at the starting value of voltage $U_0$
involves
the planar ($\theta=0$)
twisted structures of the form~\eqref{eq:clc_helix}
assuming that the azimuthal angle, $\phi_{+}$, at the top substrate
uniformly varies along the path,
whereas the angle $\phi_{-}$ is kept fixed.
Then the MEP obtained at the initial step
with $U=U_0$
is used as the initial approximation
for the MEP computed at the voltage $U=U_1=U_0+\Delta U$.
More generally, our approach can be described as
the iteration procedure where
the MEP computed at the step
with $U=U_i$
serves as the starting approximation
for the MEP evaluated at
the subsequent step with
the voltage
$U_{i+1}=U_i+\Delta U$.
Clearly, by using such a smooth continuation
of the MEPs we assume
that variations of the voltage
result in continuously dependent on $U$
deformations of the MEPs.

Figure~\ref{subfig:planar-to-tilt} shows
the energies of director configurations along the  MEPs computed at
the anchoring energy strengths $W_{\phi}=0.05$~mJ/m$^2$ and
$W_{\theta}=0.5$~mJ/m$^2$ when the voltage increases from
$U=1.37$~V to $U=1.38$~V.
The curves for the energy reveal
the presence of the pronounced peak
that increases steeply as the voltage approaches
$U=1.38$~V.
Figure~\ref{fig:saddle-images}
demonstrates how out-of-plane deformations
of the saddle-point structures
develop in the course of the transition
that might be called
the \textit{director slippage-to-anchoring breaking transition}.
The profiles of
the tilt angle for these structures are presented in
Fig.~\ref{fig:theta}.

An alternative initial approximation is
the so-called anchoring breaking approximation
used in Ref.~\cite{Kiselev:pre:2019}.
In this approximation
the uniform twist
from $\phi_{-}=0$ to $\phi_{+}$
is superimposed by the out-of-plane director deformation
with the tilt angle $\theta$ 
varying from $\theta_{-}=0$ to $\theta_{+}$.
In the first half of the path
the tilt angle $\theta_{+}=0$ at the top substrate,
increases from zero up to $\theta_{+}\approx\pi/2$.
Then in the second half of the path
the angle $\theta_{+}$ goes back to
its initial value $\theta_{+}=0$
at the final state. 
So, the anchoring breaking initial guess for the MEP
assumes that, for the transition state (saddle-point state),
the CLC director at the top substrate
is nearly normal to the bounding surface.

From the above discussion,
at voltages higher than approximately $1.38$~V,
the saddle point structure is
characterized by pronounced tilt (out-of-plane) deformations.
So, the high voltage regime,
the director slippage and anchoring breaking
approximations will produce identical results for the MEPs.
These MEPs can now be smoothly continued
to the region of low voltages.
Figure~\ref{subfig:tilt-to-planar}
presents the curves for the energies of director structures
along the
MEPs evaluated at when the voltage decreases from
$U=1.122$~V to $U=1.119$~V.
It can be seen that
behaviour of
the peak corresponding to the energy of the saddle point structure
bears a close resemblance to the
director slippage-to-anchoring breaking transition
that takes place at higher voltages
(see Fig.~\ref{subfig:planar-to-tilt}).
The only difference is the direction of the voltage changes.
So, the energy curves for the MEPs continued from high voltages
to low ones
show the behaviour that can be interpreted as the \textit{anchoring
  breaking-to-director slippage transition}.

 The energy barriers computed from
a smooth continuation of the MEPs
when an increase in applied voltage
is followed by a decrease
back to low voltages
are depicted in Fig.~\ref{fig:hysteresis}.
Clearly, the curves form the hysteresis loop.
Figure~\ref{fig:hysteresis}
also demostrates what happen
to the hysteresis loop
when the polar anchoring strength is weakened.

%%%%%%%%%%%%%%%%%%%%%%%%%%%%
\section{Conclusion}
\label{sec:conclusion}
%%%%%%%%%%%%%%%%%%%%%%%%%%%%

In this paper, we have studied
electric field dependence of
the minimum energy paths (MEPs) for
the transition between
helical structures
from different metastable branches
in the chiral nematic liquid crystal cell.
Such pathways connect
the branches
that are parametrized by the free twisting number and are
labeled by half-turn numbers.
We have examined local stability
of the twisted states forming the branches and have deduced
the analytical expressions
describing  dependence of the Fr\'eedericksz thresholds
on the free twisting number.
Our analysis have shown that
the stability conditions with respect to out-of-plane fluctuations
impose additional restrictions on the values of the free twisting number
of metastable helical states.
Such restrictions may considerably affect
mechanisms of the pitch transitions.

The energy of the saddle-point structure (transition state) of the paths gives
the energy barrier separating the metastable states.
Therefore, the MEP and its saddle points characterize
the mechanism (scenario) of the transition.
We have employed
the geodesic nudged elastic band
(GNEB) method as a computational procedure
to evaluate the MEPs. This method
requires an initial guess for the path
and various starting approximations can
generally produce different MEPs.

In our approach,
we have evaluated
two parts of
the electric field dependence of the MEPs
as smooth continuations
of the starting paths
in the low  and high voltage regions, respectively.
At low voltages,
the initial path was obtained
using
the director slippage approximation.
It turned out that,
when the voltage increases,
the energy of the saddle point structure
steeply grows in the interval
close to the critical voltage
of
the director slippage-to-anchoring breaking transition,
$U\equiv U_{c\uparrow}\approx 1.38$~V.
This is the voltage where
the nearly planar transition state
becomes tilted and
assumes noticeable out-of-plane director deformation.
In other words, this implies
that the director slippage scenario
transforms into the anchoring breaking scenario.

Similarly, when the MEP is smoothly continued from
the high voltage region dominated by
the anchoring breaking mechanism
to the region of low voltages
the director slippage scenario comes into play
when the voltage is smaller than
its critical value
$U_{c\downarrow}\approx 1.12$~V
associated with
the anchoring breaking-to-director slippage transition.
Since $U_{c\downarrow}<U_{c\uparrow}$,
the voltage dependence
of the energy barrier
is characterized by the hysteresis loop
shown in Fig.~\ref{fig:hysteresis}.

Interestingly,
the critical field
$U_{c\downarrow}$ is lower than
the Fr\'eedericksz thresholds
of both the initial and final states of the MEPs,
$U_{\ind{th}}^{(1)}\approx 1.3$~V
and
$U_{\ind{th}}^{(2)}\approx 1.67$~V.
It means that
the anchoring breaking scenario
with
the tilted saddle point
structure may take place
even if the initial and the final states are both planar.
Similar remark applies to
the full-turn state with $k=2$
that remains undistorted
at $U_{c\uparrow}<U_{\ind{th}}^{(2)}$. 

The effect of polar anchoring 
on the form of the hysteresis loop
is illustrated in Fig.~\ref{fig:hysteresis}.
It is shown that,
when the polar anchoring strength
$W_{\theta}$
is reduced by five times,
the magnitude
of energy barriers
suffer similar reduction.
The effect responsible for such behaviour
is that
the dominating contribution to
the energy of the saddle-point structure
appears to
be governed by its surface part.

Referring to Fig.~\ref{fig:hysteresis},
it can also be seen
that the
hysteresis loop widens
when the polar anchoring weakens.
Such widening can be attributed to
the anchoring induced lowering of
the Fr\'eedericksz thresholds
for the helical states.

% critical voltage
% associated with
% the anchoring breaking-to-director slippage transition,
% $U_{c\downarrow}$,
% decreases as $W_{\theta}$ is lowered,
% whereas the threshold of
% the director slippage-to-anchoring breaking transition,
% $U_{c\uparrow}$, remains essentially intact.

 \begin{acknowledgments}
 This work was supported by
 the Russian Science Foundation
 under grant 19-42-06302.  
 \end{acknowledgments}
 
%\bibliographystyle{apsrev}

%\bibliography{optics,polymer,scatter,lc,quant,hk,flc,qft,math,my_papers}
%\bibliography{tenishch}


\end{document}